\documentclass[twocolumn,showpacs,amsmath,amssymb,pra,floatfix]{revtex4}

\usepackage{graphicx}
\usepackage{dcolumn}
\usepackage{bm}
\usepackage{bbm}
\usepackage{multirow,amssymb,amsbsy,amsmath}
\usepackage{stmaryrd}

\newcommand{\I}{\textup{i}}
\newcommand{\E}{\textup{e}}

\newcommand{\ddim}{\udelta\kern0.1em}

\newcommand{\beikonst}[2]{\left( #1 \right)_{\kern-0.2em #2}}
\newcommand{\tr}[2][]{\text{Tr}_{#1}\left\{#2\right\}}

\newcommand*{\bra}[1]{\mathopen{\langle}#1\mathclose{|}}
\newcommand*{\ket}[1]{\mathopen{|}#1\mathclose{\rangle}}
\newcommand{\braket}[2]{\mathopen{\langle}#1\mathclose{|}#2\mathclose{\rangle}}

\newcommand{\dop}{\hat{\rho}_A}

\newcommand{\rabi}[1]{\Omega_{#1}}
\newcommand{\nbar}{\bar{n}}

\renewcommand{\Re}{\textup{Re\,}}
\renewcommand{\Im}{\textup{Im\,}}

\begin{document}
\preprint{APS/123-QED}

%
%
\title{Cavity-induced temperature control of a two-level system}

\author{Hendrik Weimer}%
\affiliation{Institute of Theoretical Physics I, University of Stuttgart, %
             Pfaffenwaldring 57, 70550 Stuttgart, Germany}%
\email{hweimer@itp1.uni-stuttgart.de}%
\author{G\"unter Mahler}%
\affiliation{Institute of Theoretical Physics I, University of Stuttgart, %
             Pfaffenwaldring 57, 70550 Stuttgart, Germany}%

\date{\today}%

\begin{abstract}
  We consider a two-level atom interacting with a single mode of the
  electromagnetic field in a cavity within the Jaynes-Cummings model.
  Initially, the atom is thermal while the cavity is in a coherent
  state. The atom interacts with the cavity field for a fixed time.
  After removing the atom from the cavity and applying a laser pulse
  the atom will be in a thermal state again. Depending on the
  interaction time with the cavity field the final temperature can be
  varied over a large range. We discuss how this method can be used to
  cool the internal degrees of freedom of atoms and create heat baths
  suitable for studying thermodynamics at the nanoscale.
\end{abstract}


\maketitle

%
%


The Jaynes-Cummings model \cite{Jaynes1963} (JCM) is a simple but
powerful model describing the interaction between a two-level atom and
a single mode of the radiation field. While being exactly solvable it
offers a large variety of genuinely quantum phenomena like collapses
and revivals in the inversion of the atom \cite{Yoo1985,Shore1993},
which have been observed experimentally as well
\cite{Rempe1987,Brune1996}.

For a field prepared in a coherent state, the state of the atom will
be almost pure at half of the revival time if the atom is initially in
a pure state \cite{Gea-Banacloche1990,Phoenix1991}. However, a more
realistic model would involve a thermal initial state for the atom.
The thermal contribution to the initial state of the field may be
neglected as long as the number of coherent photons is sufficiently
larger than the number of thermal photons \cite{Satyanarayana1992}.
Using thermal states allows for an investigation of the thermal
properties of the JCM, i.e., its applicability for problems like the
initial state preparation in quantum computing \cite{DiVincenzo2000},
cooling of atoms \cite{Chu1998,Cohen-Tannoudji1998,Phillips1998}, or
implementation of quantum thermodynamic machines \cite{Gemmer2004}.

\begin{figure}
  \includegraphics{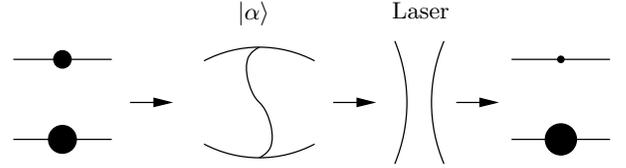}
  \caption{Outline of the procedure: A two-level atom in a thermal
    state (occupation probabilities represented by black dots)
    interacts with a cavity prepared in a coherent state
    $\ket{\alpha}$. After a time $t$ the atom leaves the cavity. A
    laser pulse is applied to the system, resulting in a thermal state
    with a different temperature.}
  \label{fig:proc}
\end{figure}

In the following we will discuss a model where an atom in a thermal
state enters a cavity prepared in a coherent state. By obtaining a
closed form for the reduced density matrix for the atom we will show
that after the collapse the state of the atom is independent of its
initial state. After a fixed interaction time the atom is taken to
leave the cavity and to interact with a laser field, which is treated
as a semi-classical driver. For an appropriate laser field the final
state will be thermal as well. Depending on the interaction time with
the cavity the final temperature can be varied over a large range,
leading to cooling or heating of the atom.  We will present an
expression for the minimum and maximum temperature that can be
achieved. Finally, we will discuss applications of the method to
cooling of the internal degrees of freedom of atoms and creating heat
baths suitable for studying thermodynamics at the nanoscale. The whole
procedure of our proposal is depicted in Fig.~\ref{fig:proc}.


The total system is described by the Hamiltonian
\begin{equation}
  \hat{H} = \hat{H}_A + \hat{H}_F + \hat{H}_I,
\end{equation}
where the atomic Hamiltonian $\hat{H}_A$ is given by
\begin{equation}
  \hat{H}_A = \frac{\Delta E}{2}\hat{\sigma}_z,
\end{equation}
with $\Delta E$ being the energy splitting. The field Hamiltonian
$\hat{H}_F$ is
\begin{equation}
  \hat{H}_F = \hbar\omega\left(\hat{a}^{\dagger}\hat{a} + \frac{1}{2}\right),
\end{equation}
with $\omega$ being the frequency of the single mode and $\hat{a}$
being the annihilation operator of the field. Being in resonance, we
have $\omega = \Delta E/\hbar$. Using the electric dipole and rotating
wave approximation the JCM interaction Hamiltonian is given by
\begin{equation}
  \label{eq:JCM}
  \hat{H}_I = g\hat{\sigma}^+\hat{a} + g^*\hat{\sigma}^-\hat{a}^{\dagger},
\end{equation}
where $g$ is the coupling constant for the atom-field interaction and
$\hat{\sigma}^{\pm}$ are the atomic transition operators.
Particularly, the coupling constant is given by
\begin{equation}
  \label{eq:g}
  g = d\sqrt{\frac{\omega}{\hbar\varepsilon_0 V}},
\end{equation}
where $d$ is the atomic electric dipole matrix element and $V$ is the
mode volume.
  

We restrict ourselves to the field being initially in a coherent state
$\ket{\alpha}$ and the atom being in a thermal state described by the
density operator
\begin{equation}
  \dop(0) = Z^{-1} \exp(-\beta\hat{H}_A) \equiv p_e(0) \ket{e}\bra{e}+[1-p_e(0)]\ket{g}\bra{g},
\end{equation}
with $Z$ being the partition function, $\beta$ the inverse
temperature, $p_e$ the probability to find the atom in its excited
state $\ket{e}$, and $\ket{g}$ its ground state.


The time evolution of the full system is then given by
\begin{eqnarray}
  \hat{\rho}(t) &=& p_e(0)\hat{U}\ket{e,\alpha}\bra{e,\alpha}\hat{U}^{\dagger} + [1-p_e(0)]\hat{U}\ket{g,\alpha}\bra{g,\alpha}\hat{U}^{\dagger}\nonumber\\
  &\equiv& p_e(0)\ket{\psi_e(t)}\bra{\psi_e(t)}+[1-p_e(0)]\ket{\psi_g(t)}\bra{\psi_g(t)},\nonumber\\
\end{eqnarray}
where $\hat{U}$ is the time evolution operator of the full system.

In order to obtain the effective time evolution for
the atom alone, the degrees of freedom corresponding to the field have
to be traced out \cite{Breuer2002}. Here, the partial trace over the
field is given by
\begin{equation}
  \dop(t) = \tr[F]{\ket{\psi(t)}\bra{\psi(t)}} = \sum\limits_n \braket{n}{\psi(t)}\braket{\psi(t)}{n}.
\end{equation}
Since $\dop$ is Hermitian and has unit trace, the atom is effectively
described by the diagonal element $\rho_{11}$ and the off-diagonal
element $\rho_{01}$.

In the following we first consider the case where the initial state is
$\ket{e,\alpha}$. Then, the full time evolution is given by (see,
e.g., \cite{Basdevant2000})
\begin{align}
  \label{eq:full}
  \ket{\psi_e(t)} = \sum\limits_n&\left(\E^{-\I\rabi{n+1}t/2}\ket{+_{n+1}}-\E^{\I\rabi{n+1}t/2}\ket{-_{n+1}}\right)\nonumber\\ & \times \frac{\E^{-|\alpha|^2/2}}{\sqrt{2}} \frac{\alpha^n}{\sqrt{n!}}\E^{-\I(n+1/2)\omega t},
\end{align}
where the $n$-photon Rabi frequency $\rabi{n} = g\sqrt{n}$ and the
$n$-photon eigenstates of the atom-field system,
\begin{equation}
  \label{eq:pm}
  \ket{\pm_n} = \frac{1}{\sqrt{2}}(\ket{g,n+1}\pm\ket{e,n}),
\end{equation}
have been used. The time evolution of the reduced density matrix
element $\rho_{11}$ before the revival time have been studied
extensively (see, e.g., \cite{Basdevant2000}) and is given by
\begin{equation}
  \label{eq:r11}
  \rho_{11}(t) = \frac{1}{2} + \frac{1}{2}\cos(2gt)\exp\left(-\frac{t^2}{\tau_C^2}\right),
\end{equation}
with $\tau_C$ being the collapse time $\tau_C = \sqrt{2}/g$.
For an atom initially in $\ket{g}$ the result is
\begin{equation}
  \label{eq:r11g}
  \rho_{11}(t) = \frac{1}{2} - \frac{1}{2}\cos(2gt)\exp\left(-\frac{t^2}{\tau_C^2}\right).
\end{equation}
Therefore, after the collapse the diagonal elements are constant and
$\rho_{ii} = 1/2$.

\begin{figure}
  \centering 
  \includegraphics{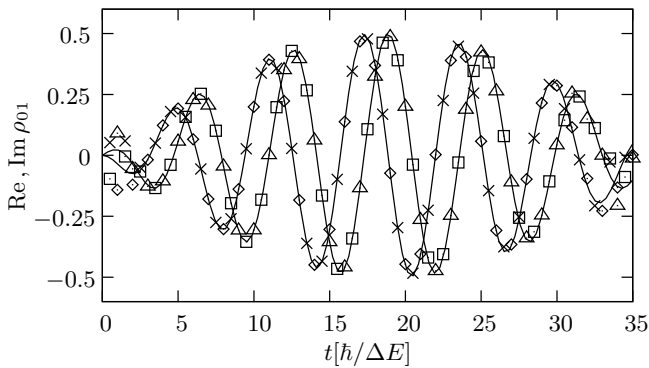}
  \caption{Comparison of the real and imaginary part of
    Eq.~(\ref{eq:r01}) (solid lines) and the solution of the full
    time-dependent Schr\"odinger equation. Initial states for the atom
    were $\ket{g}$ ($\Re \rho_{01}$: crosses, $\Im \rho_{01}$: boxes)
    and $\ket{e}$ ($\Re \rho_{01}$: diamonds, $\Im \rho_{01}$:
    triangles). ($\nbar = 36$, $g = \Delta E$, and $\phi = 0$)}
\label{fig:r01}  
\end{figure}

The off-diagonal element $\rho_{01}$ (again, first for the atom
initially in $\ket{e}$) is given by
\begin{equation}
  \label{eq:sum}
  \rho_{01}(t) = \sum\limits_n\braket{\psi_e(t)}{n,g}\braket{n,e}{\psi_e(t)}.
\end{equation}
Evaluating the summands $\rho_{01}^{(n)}$ using Eq.~(\ref{eq:full})
and Eq.~(\ref{eq:pm}) leads to
\begin{align}
  \label{eq:wahrheit}
\rho_{01}^{(n)}(t)= \I w(n) \frac{\sqrt{n}}{2\alpha^*} \E^{-\I\omega t}&\left\{\sin\left[(\rabi{n+1}+\rabi{n})\frac{t}{2}\right]\right.\nonumber\\
& \left.-\sin\left[(\rabi{n+1}-\rabi{n})\frac{t}{2}\right]\right\},
\end{align}
with $w(n)$ being the Poisson distribution. The first term inside the
square brackets oscillates at a much higher frequency than the second
and results only in a random phase, which vanishes after summation. In
the high-photon limit $\sqrt{n}$ may be approximated by (see
\cite{Gea-Banacloche1990})
\begin{equation}
  \label{eq:sqrtn}
  \sqrt{n} \approx \sqrt{\nbar} + \frac{n-\nbar}{2\sqrt{\nbar}}.
\end{equation}
Analogously, the difference of the Rabi frequencies can be expressed
as
\begin{eqnarray}
  \rabi{n+1}-\rabi{n} &=& 2g(\sqrt{n+1}-\sqrt{n})\\
\label{eq:rabi} &\approx& 2g\left(\frac{1}{2\sqrt{\nbar}}-\frac{1}{8\sqrt{\nbar^3}} - \frac{n-\nbar}{4\sqrt{\nbar^3}}\right).
\end{eqnarray}
Plugging only the leading order into Eq.~(\ref{eq:wahrheit}) and
replacing the sum in Eq.  (\ref{eq:sum}) by an integral over a
Gaussian distribution leads to
\begin{equation}
  \label{eq:r01}
  \rho_{01}(t) = -\frac{\I}{2}\exp[\I(\omega t+\phi)]\sin\frac{gt}{2\sqrt{\bar{n}}},
\end{equation}
where $\phi$ is the initial phase of the radiation field. Using the
same approximations for the atom initially in its ground state yields
the same result for $\rho_{01}(t)$. Therefore, after the collapse the
atom evolves totally independent from its initial state. A comparison
of Eq.~(\ref{eq:r01}) with the numerical solution of the full
time-dependent Schr\"odinger equation is shown in Fig.~\ref{fig:r01}.
Apart from the collapse and revival phase there is excellent
agreement. This further shows that the random phase approximation
applied to Eq.~(\ref{eq:wahrheit}) was perfectly justified.

\begin{figure}
  \centering 
  \includegraphics[width=4cm]{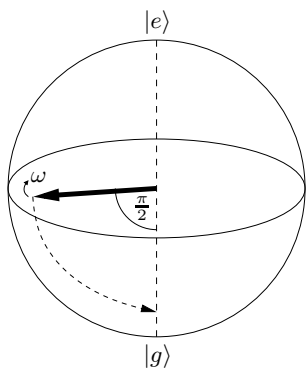}
  \caption{Illustration of the $\pi/2$ pulse acting on the Bloch
    vector of the atom.}
\label{fig:bloch}  
\end{figure}

Since the diagonal elements of $\dop$ are both at $\frac{1}{2}$ the
Bloch vector only moves within $x-y$ plane of the Bloch
sphere. Furthermore, the phase oscillates at $\omega$, i.e., in the
rotating frame only the initial phase $\phi$ is relevant [see
Eq.~(\ref{eq:r01})]. Therefore, in order to obtain a thermal state one
always has to apply a $\pi/2$ pulse to the system (see
Fig.~\ref{fig:bloch}).  Since the pulse diagonalizes $\dop$, the
probability to find the atom in its excited state after the pulse
$p_e(t)$ is given by the smallest eigenvalue of $\dop$. Computation of
$p_e(t)$ yields
\begin{equation}
  \label{eq:pe}
  p_e(t) = \frac{1}{2}\left(1-\sin\frac{gt}{2\sqrt{\bar{n}}}\right).
\end{equation}
This can also be expressed as a temperature using
\begin{equation}
  \label{eq:T}
  T = -\frac{\Delta E}{k_B\log\left(\frac{p_e}{1-p_e}\right)}.
\end{equation}

This temperature should be considered as a parameter characterizing
the mixedness of the output state rather then as an indication for a
stable thermal state proper. Quantum objects prepared like this (or
ensembles thereof) would then constitute resources for further
applications like quantum information processing.

\begin{figure}
  \centering 
  \includegraphics{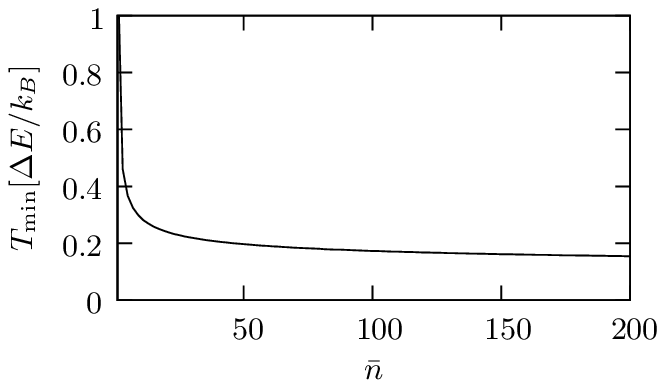}
  \caption{Minimum temperature $T_{\mathrm{min}}$ over average photon number
    $\nbar$.}
\label{fig:Tmin}  
\end{figure}

Equation (\ref{eq:pe}) suggests that at half of the revival time the
atom will be in its ground state (i.e., $T=0$). However, this minimum
temperature would only be reached for infinitely large $\bar{n}$, for
which it would take an infinitely long time to reach this state. In
order to determine the actual minimum temperature a correction for
finite $\nbar$ is required. A correction to Eq. (\ref{eq:pe}) can be
obtained by including the next order in Eq. (\ref{eq:rabi}).  Close to
half of the revival time the sine in Eq. (\ref{eq:wahrheit}) is near
its maximum and can be approximated by a second order Taylor
expansion, which leads to a final result of
\begin{equation}
  p_e\left(\frac{\tau_R}{2}\right) = \frac{\pi^2}{32\bar{n}}.
\end{equation}
Using the next order in Eq. (\ref{eq:sqrtn}) as well leads to an
additional correction in $O(1/\nbar^2)$. Putting this $p_e$ into
Eq. (\ref{eq:T}) gives the minimum temperature
$T_{\mathrm{min}}(\nbar)$ as shown in
Fig.~\ref{fig:Tmin}. Temperatures as low as $0.2\,\Delta E/k_B$ can be
obtained, which correspond to an occupation probability of the excited
state of the order of $10^{-3}$.

\begin{figure}
  \centering 
  \includegraphics{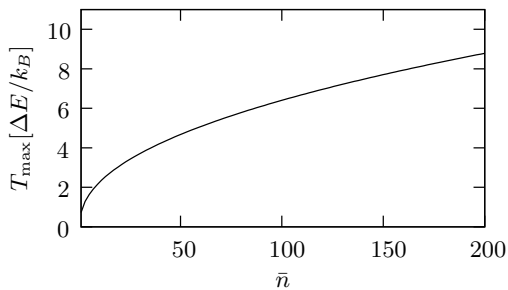}
  \caption{Maximum temperature $T_{\mathrm{max}}$ over average photon number
    $\nbar$.}
\label{fig:Tmax}  
\end{figure}

In order to determine the maximum temperature that can be reached we
require that the collapse must have taken place [i.e,. the difference
in the occupation probabilities Eqs.~(\ref{eq:r11}) and
(\ref{eq:r11g}) is negligible compared to the difference induced by
the laser].  Requiring the former to be smaller by a factor of $10$,
this can be expressed as
\begin{equation}
  10 \cos(2gt)\exp\left(-\frac{t^2}{\tau_C^2}\right) = \sin\frac{gt}{2\sqrt{\bar{n}}}.
\end{equation}
The cosine on the left hand side may be replaced by unity without
violating the above requirement. For large $\nbar$ the right hand side
can be approximated linearly in $t$, resulting in
\begin{equation}
  10 \exp\left(-\frac{t^2}{\tau_C^2}\right) = \frac{gt}{2\sqrt{\bar{n}}}.
\end{equation}
Solving for the appropriate cavity interaction time $t$ and using Eqs.
(\ref{eq:pe}) and (\ref{eq:T}) leads to a maximum temperature
$T_{\mathrm{max}}$ of
\begin{equation}
  T_{\mathrm{max}} = \frac{\Delta E}{k_B\log \frac{4\sqrt{\nbar}+\sqrt{W(400\nbar)}}{4\sqrt{\nbar}-\sqrt{W(400\nbar)}}},
\end{equation}
where $W(\cdot)$ denotes the Lambert $W$ function, i.e., the inverse
function of $f(x)=x\E^x$. Figure~\ref{fig:Tmax} shows the dependence
of $T_{\mathrm{max}}$ on $\nbar$.


These results show that the temperature can be tuned over a large
range, which depends only on the average photon number $\nbar$, the
coupling time $t$, and the energy splitting $\Delta E$.  However,
there are some other applications for this procedure, which are
realizable within present experimental setups. A rather obvious one is
the cooling of the internal degrees of freedom of atoms.  However, an
implementation using a cavity would be extremely difficult as the
frequencies relevant for cooling are in the MHz range, where the
coupling constant $g$ is much too small to observe any effects [due to
the $\omega$ dependence in Eq.  (\ref{eq:g})]. A much more promising
implementation could be realized using circuit quantum electrodynamics
(QED) \cite{Blais2004,Wallraff2004}, in which the atom is replaced by
a Cooper-pair box and the cavity is implemented by a one-dimensional
resonator. There, the coupling constant is sufficiently large even in
the relevant frequency range. Although dephasing plays an important
role in circuit QED, the dephasing time is of the order of several
hundred nanoseconds \cite{Blais2007}, while the required time for
reaching the minimum temperature is about one order of magnitude
smaller for experimentally feasible parameters. Using our procedure
thus might lead to lower temperatures than currently employed
techniques. Besides circuit QED, other implementations involving a
Jaynes-Cummings Hamiltonian with a tunable coupling constant may prove
useful as well.

Another interesting application of this procedure could be the
realization of tiny local baths. Local baths are an important
ingredient in nonequilibrium quantum thermodynamics \cite{Gemmer2004},
where it is necessary to create and control a temperature gradient on
a nanoscopic scale.  This could be used to investigate transport
behavior \cite{Saito2000,Michel2003} or quantum thermodynamic machines
\cite{Henrich2006}. Using our framework to repeatedly set a
temperature of a single two-level system could act as such a local
bath as long as the cavity is reset after each step and the
temperature control happens on a much smaller timescale than the other
processes within the system (i.e., strong bath coupling).


In summary, we have shown that the temperature of a two-level atom
could be efficiently controlled via a resonant interaction with a
cavity.  Depending only on the interaction time with the cavity, it
should be possible to tune the final temperature over a large range.
The expression for the reduced density matrix of the atom has been
obtained in the high photon limit using a systematic series expansion
and has been verified by comparison with the solution of the full
time-dependent Schr\"odinger equation. Besides temperature control our
procedure may prove useful for cooling various microscopic systems or
realizing local baths in nanothermodynamics.

We thank T.\ Pfau, M.\ Michel, M.\ Henrich, F.\ Rempp, G.\ Reuther,
H.\ Schmidt, H.\ Schr\"oder, and M. Youssef for fruitful discussions.


\end{document}